\documentclass[pre,twocolumn,aps,eqsecnum]{revtex4}
\usepackage{epsfig}

\begin{document}

\title{Glass Dynamics at High Strain Rates}

\author{J.S. Langer}
\affiliation{Department of Physics, University of California, Santa Barbara, CA  93106-9530}

\author{Takeshi Egami}
\affiliation{
Joint Institute for Neutron Sciences, Department of Materials Science and Engineering, Department of Physics and Astronomy, University of Tennessee, Knoxville, TN 37996-1580; and 
Oak Ridge National Laboratory, Oak Ridge, TN 37831}

\date{\today}

\begin{abstract}
We present a shear-transformation-zone (STZ) theoretical analysis of molecular-dynamics simulations of a rapidly sheared metallic glass.  These simulations are especially revealing because, although they are limited to high strain rates, they span temperatures ranging from well below to well above the glass transition.  With one important discrepancy, the simplified STZ theory used here reproduces the simulation data, including the way in which those data can be made to collapse approximately onto simple curves by a scaling transformation.  The STZ analysis implies that the system's behavior at high strain rates is controlled primarily by effective-temperature thermodynamics, as opposed to system-specific details of the molecular interactions.  The discrepancy between theory and simulations occurs at the lower strain rates for temperatures near the glass transition.  We argue that this discrepancy can be resolved by the same multi-species generalization of STZ theory that has been proposed recently for understanding frequency-dependent viscoelastic responses, Stokes-Einstein violations, and stretched-exponential relaxation in equilibrated glassy materials. 

\end{abstract}
\maketitle

\section{Introduction}
\label{intro}
A remarkable scaling property of viscosity as a function of shear stress has been observed by by Olsson and Teitel  \cite{OLSSEN-TEITEL-07} in simulations of a strongly overdamped, athermal, two-dimensional, amorphous system near its jamming transition.  More recently, an approximately similar scaling behavior has been seen by Guan, Chen and Egami \cite{GCE-10} (GCE) in molecular dynamics simulations of a rapidly sheared, three dimensional metallic glass that undergoes a thermodynamic glass transition.  These results suggest that there may be a useful, minimal description of the underlying physics for the seemingly complex dynamics of flow in glass-forming fluids, and that some scaling properties of a zero-temperature jamming transition may persist in rapidly deforming systems far from jamming conditions. 

The distinguishing feature of the GCE simulations is that they cover an unusually wide range of system parameters.  They are carried out at kinetic temperatures ranging from $100\,K$ to $1100\,K$, i.e., from well below to well above an  apparent glass transition.  Importantly, they are carried out at strain rates from $10^8\,s^{-1}$ to $10^{12}\,s^{-1}$, all well above values ordinarily accessible in the laboratory. By driving their systems at large strain rates, GCE achieve steady-state, nonequilibrium shear flows at temperatures far below the glass transition.  Under stress, molecular rearrangements become faster and the internal disorder increases, while the kinetic temperatures remain low. GCE are not probing truly glassy behavior in this way; however, they are discovering important features of glass dynamics by looking from this point of view.

In this paper, we explore the implications of the GCE results by comparing them with the predictions of a shear-transformation-zone (STZ) theory \cite{FL-98,JSL-08,FL-11}.  This comparison is timely because Voigtmann \cite{VOIGTMANN-11} recently has published an analysis of the same data using an extended version of mode-coupling theory (MCT) \cite{GOTZE-91,GOTZE-92,BRADER-et-al-09}.  These two theories of amorphous plasticity would seem at first sight to have non-overlapping regions of validity. 

MCT starts with a liquidlike description of a many-body system.  Its strength is that the coupling terms that emerge from its approximate closure of the many-body equations of motion can be evaluated in terms of observable structure factors, thus basing the theory directly upon microscopic dynamics.  Its weakness, however, is that this closure approximation is accurate only if the density fluctuations are distributed Gaussianly.  Therefore, MCT is effectively a mean-field theory in which motions are determined by averaging over large numbers of weakly correlated events.  Such an approximation becomes qualitatively incorrect at low temperatures, where the flow is governed by sporadic, thermally activated events that are rare but intense \cite{CATES-PTP-10}. Voigtmann's results may imply that MCT retains some phenomenological validity for flowing states in the low-temperature regime; but they cannot in principle have the same fundamental validity that MCT can claim at higher temperatures. 

In contrast, the STZ theory starts with a solidlike picture in which thermally activated, localized flow defects play the central role in controlling plastic deformation. It predicts that, at low temperatures, the yielding transition as a function of stress is an exchange of dynamic stability between jammed and flowing states; and it describes the mechanical behavior of the system on both sides of that transition.  When combined with an equation of motion for the effective disorder temperature of the configurational (i.e. structural) degrees of freedom, the STZ theory successfully predicts shear-banding instabilities in agreement with simulations \cite{SHIetal07,MANNINGetal-SHEARBANDS-07}  and, as will be seen here, describes a transition to liquidlike behavior at large stresses.  It is in the latter regime, however, that the STZ theory encounters a weakness complementary to that of MCT.  The highest GCE strain rates are comparable to inverse molecular relaxation times, where the system must become a true fluid, and where the flow defects are no longer dilute and dynamically independent of each other.  Nevertheless, like MCT at low temperatures, the STZ theory seems to retain at least a phenomenological validity up to quite high strain rates.  Why does this happen?  Is there a region of overlapping validity of the MCT and STZ theories?  Is this theoretical agreement related to the approximate scaling behavior observed by GCE?  

Our tentative answer to these questions is that, to a first approximation, our results are sensitive to only a few features of the molecular interactions. The observed behavior is governed largely by general thermodynamic principles supplemented by dimensional analysis.  The relation between stress and strain rate in the rapid-deformation limit is determined almost entirely by the density of STZ's, which is determined by the effective disorder temperature.  The details of specific mechanisms, such as the thermal activation rate for internal STZ transitions, are almost irrelevant in most, but not all, of the regime of interest.  

Interestingly, the agreement between the STZ theory and the GCE simulations fails for the smaller strain rates at temperatures near the glass transition -- for reasons that we think we understand.  All of the analysis presented here is based on a single-species STZ model, which we know to be incorrect for weakly perturbed, well equilibrated, glass forming systems at low temperatures.  Under the latter conditions, a statistical argument predicts that STZ's with a wide range of internal transition rates become thermodynamically stable, and that the STZ's at the slow end of this range control the viscous response \cite{BL-11}.   This distribution of transition rates accounts accurately for the very broad response peaks observed in oscillatory viscoelastic measurements, and for related experimental observations such as Stokes-Einstein violations and stretched-exponential relaxation \cite{JSL-11}.  Indeed, the success of the STZ theory in predicting these intrinsically glassy, near-equilibrium phenomena gives us confidence in the present attempt to extend the analysis to strongly driven situations; but it also tells us where to expect difficulties.

We start, in Section \ref{STZsummary}, by presenting the steady-state STZ equations of motion in a more general and compact form than the one that has appeared in previous publications \cite{FL-98,JSL-08,FL-11}. Then, in Sec.~\ref{chi}, we summarize the primarily kinematic arguments that lead to our equation of motion for the effective temperature.  In Sec.~\ref{parameters}, we describe the ways in which we use the simulation data to evaluate the theoretical parameters. Section \ref{analysis} is devoted to our interpretation of the GCE scaling analysis.  We conclude in Sec.~\ref{conclusions} with a discussion of open questions. 

\section{STZ Dynamics}
\label{STZsummary}

The basic premise of the STZ theory is that, in a closely packed, amorphous material, deformation is enabled by a fluctuating population of rare, localized, two-state, flow defects, i.e. STZ's. Let $n_{\pm}$ denote the number densities of STZ's oriented parallel and antiparallel to a deviatoric shear stress $s$. These densities satisfy a master equation of the form
\begin{eqnarray}
\label{ndot}
\nonumber
\tau_0\,\dot n_{\pm} &=& R(\pm s)\,n_{\mp} - R(\mp s)\,n_{\pm}\cr \\&+& \tilde\Gamma(s,\chi)\,\left[n_{\infty}(\chi) - n_{\pm}\right].
\end{eqnarray} 
On the left-hand side, the factor $\tau_0$ is a microscopic time scale, of the order of picoseconds for metallic glasses. On the right-hand side, the first terms containing the factors $R(\pm s)$ are the rates of forward and backward STZ transitions (in units of $\tau_0^{-1}$).  In general, these rate factors are functions of both the ordinary and effective temperatures.  They  determine the plastic strain rate $\dot\gamma^{pl}$, which we write in the dimensionless form:
\begin{equation}
\label{qeqn}
q \equiv \tau_0\,\dot\gamma^{pl} = \epsilon_0\,v_0\,\left[R(+s)\,n_- -R(- s)\,n_+\right],
\end{equation}
where $v_0$ is the average volume per molecule, and $\epsilon_0$ is a dimensionless constant of the order of unity.

The second terms on the right-hand side of Eq.(\ref{ndot}) are the rates of STZ creation and annihilation.
These are fluctuation-activated processes, expressed in the form of a detailed-balance relation in which the steady-state  STZ density is determined by the effective temperature $\chi$:
\begin{equation}
\label{neq}
n_{\infty}(\chi) = {1\over v_0}\,e^{-1/\chi}.
\end{equation}
Here, $\chi$ is written in units of the STZ formation energy $e_Z$, i.e. $\chi = k_B\,T_{e\!f\!f}/e_Z$.  To a first approximation, $e_Z$ is the energy required to create the amorphous analog of a vacancy-interstitial pair.  Note that the effective temperature introduced here is the one that is defined thermodynamically in \cite{BLII-III-09}.  We presume that it is the same as the effective temperature determined by a fluctuation-dissipation relation, as described in \cite{CUGLIANDOLO-11} and references cited there; but we know of no rigorous proof of that assertion. 

$\tilde\Gamma(s,\chi)/\tau_0$ is an attempt frequency consisting of additive thermal and mechanical parts: 
\begin{equation}
\label{tildeGamma}
\tilde\Gamma(s,\chi) = \rho + \Gamma(s,\chi).
\end{equation}
The term $\rho = \rho(\theta)$, where $\theta = k_B\,T/e_Z$, is best understood as a dimensionless, thermal noise strength. It is a super-Arrhenius factor whose strong temperature dependence governs the equilibrium glass transition, and for which one of us (JSL) has proposed the outlines of a first-principles derivation.\cite{Langer-PRE-06,Langer-PRL-06} We expect that $\rho(\theta)$ is of the order of unity at high temperatures, and that it decreases rapidly toward zero as $\theta$ falls through the glass temperature. In the absence of external driving, $\rho(\theta)$ controls the rate at which the system undergoes structural aging.  In much of the literature, the quantity $\rho(\theta)\,\exp\,(-1/\theta)\,/\tau_0$ is defined to be the $\alpha$ relaxation rate $\tau_{\alpha}^{-1}$. However, as pointed out in \cite{JSL-11}, this definition is not the same as the conventional assumption that $\tau_{\alpha}$ is directly proportional to the viscosity.  

In analogy to $\rho(\theta)/\tau_0$, the quantity $\Gamma(s,\chi)/\tau_0$ is the contribution to the attempt frequency in Eq.(\ref{tildeGamma}) due to mechanically generated noise.  As in earlier papers \cite{FL-11,LP03,PECHENIK-05}, we assume that $\Gamma(s,\chi)/\tau_0$ is proportional to the rate of entropy production.  In steady flow, all of the work done on the system is dissipated as heat; therefore, the rate of energy dissipation per unit volume is $2\,\dot\gamma^{pl}\,s$.  To convert this rate to a noise strength with dimensions of inverse time, we multiply by the volume per noise source, i.e. the volume per STZ, $v_0\,\exp\,(1/\chi)$, and divide by an energy conveniently written in the form $2\,\epsilon_0\,s_0\,v_0$.  The fact that the stress $s_0$, introduced here for dimensional reasons, turns out to be the low-temperature yield stress has been one of the more interesting surprises in this theory. The resulting formula for the noise strength is
\begin{equation}
\label{Gamma}
\Gamma(s,\chi) = {q\,s\over \epsilon_0\,s_0}\,e^{1/\chi}.
\end{equation}  

The GCE simulations involve only steady-state deformations; thus all of the information that we need is contained in the stationary solutions of Eq.(\ref{ndot}), obtained by setting $\dot n_{\pm} = 0$ and solving for the $n_{\pm}$.  The resulting expression for the strain rate in Eq.(\ref{qeqn}) is
\begin{equation}
\label{qeqn1}
q = \epsilon_0\,\tilde\Gamma(s,\chi)\,e^{-1/\chi}\,\left[{2\,{\cal C}(s)\over 2\,{\cal C}(s) +\tilde\Gamma(s,\chi)}\right]\,{\cal T}(s,\chi),
\end{equation}
where
\begin{equation}
 {\cal C}(s) = {1\over 2}\,[R(+s) + R(-s)],
\end{equation}
and 
\begin{equation}
\label{calTdef}
 {\cal T}(s,\chi) = {R(+s) - R(-s)\over R(+s) + R(-s)}.
\end{equation}
According to Eq.(\ref{calTdef}), ${\cal T}(s,\chi)$ is the bias between forward and backward transitions. The second law of thermodynamics (see \cite{BLII-III-09}) requires that 
\begin{equation}
\label{calT}
{\cal T}(s,\chi) = \tanh\,\left({v_0\,s\over e_Z\,\chi}\right). 
\end{equation}

With Eq.(\ref{Gamma}), Eq.(\ref{qeqn1}) becomes a quadratic equation that can be solved for $q$ or, more conveniently, for $\Gamma(s,\chi)$.  The result is
\begin{equation}
\label{Gamma1}
\Gamma(s,\chi) = {1\over 2}\,{\cal Q}(s,\chi)+ {1\over 2}\,\sqrt{{\cal Q}(s,\chi)^2 + 4\,\rho(\theta)\,{\cal Q}(s,\chi)},~~~~~
\end{equation} 
where
\begin{equation}
{\cal Q}(s,\chi) = 2\,{\cal C}(s)\,{\cal T}(s,\chi)\,{s\over s_0}- 2\,{\cal C}(s) - \rho(\theta).
\end{equation} 
Then, knowing $\Gamma(s,\chi)$, we can use Eq.(\ref{Gamma}) to write: 
\begin{equation}
\label{q-sol}
q = {\epsilon_0\,s_0\over s}\,e^{-1/\chi}\,\Gamma(s,\chi).
\end{equation} 
Note that $\Gamma(s,\chi)$ is a non-negative, symmetric function of $s$ that vanishes like $s^2$ at $s = 0$.  

In the athermal limit where $\rho = 0$, both $\Gamma(s,\chi)$ and $q$ vanish for $s < s_y$ (the jammed state); whereas, for $s > s_y$ (the flowing state),
\begin{equation}
\label{qeqn2}
q = 2\,\epsilon_0 \,e^{-1/\chi}\,{\cal C}(s) \left[{\cal T}(s,\chi)-{s_0\over s}\right].
\end{equation}
The yield stress $s_y$ is the solution of 
\begin{equation}
s_y\,{\cal T}(s_y,\chi) = s_0.
\end{equation}
Therefore, $s_y \cong s_0$ when $s_y$ is large and ${\cal T}(s_y,\chi) \cong 1$.  For small but nonzero $\rho$, according to Eq.(\ref{Gamma1}), both $\Gamma(s,\chi)$ and $q$ make smooth transitions near $s = s_y$ between viscous and flowing states.

We also need to examine the viscous limit at small $q$, where $\Gamma(s,\chi) \ll \rho$ and $\chi \approx \theta$.  Keeping only the linear term in the relation between $q$ and $s$, we find
\begin{equation}
\label{eta}
q = \epsilon_0\,\rho(\theta)\,e^{-1/\theta}\,\left({2\,{\cal C}(0)\over 2\,{\cal C}(0) + \rho(\theta)}\right)\,{v_0\,s\over e_Z\,\theta}.
\end{equation}
The Newtonian linear viscosity, for vanishingly small $q$, and expressed here in units of stress, is $\eta_N = s/q$. Note that this formula produces the conventional result in which $\eta_N \propto \tau_{\alpha}$ multiplied by a slowly varying function of $\theta$.  It is this result that, according to  \cite{JSL-11},  substantially underestimates $\eta_N$ at low temperatures and, therefore, is responsible for the important discrepancy between this theory and the GCE simulations.  

${\cal C}(s)$ is necessarily a symmetric function of $s$. We write it in the form:
\begin{equation}
\label{calC}
{\cal C}(s)= \exp\,\left[-{\theta_E\over \theta}\,e^{-\,s^2/2\,s_E^2}\right]\,\left[1 + \left({s\over s_B}\right)^2\right]^{1/4}.
\end{equation} 
On the right-hand side, the first factor in square brackets is a thermally activated (Eyring-like) rate in which a barrier of height $\theta_E$ is reduced quadratically by the applied stress via a Gaussian factor, thus preserving the required symmetry. In what follows, we find that we can approximate this term by unity. The second factor is similar to, but not strictly the same as a Bagnold rate, proportional to the square root of the stress at large $s$ for dimensional reasons.  It does seem to be significant at low temperatures and high strain rates.  The only quantity that can play the role of $s_B$ in this formula is the pressure, which may be roughly proportional to the temperature.  Thus, we write $s_B = (T/T_B)$ GPa, where the ``Bagnold temperature'' $T_B$ is a constant. 

\section{Effective-Temperature Kinematics}
\label{chi}

Turn now to the effective temperature $\chi$.  In general, $\chi$ is the temperature of the configurational degrees of freedom of the system, thus it characterizes the system's state of structural disorder. Its role here is primarily to determine the density of STZ's, as indicated by its appearance in the Boltzmann factor in Eq.(\ref{neq}).  We propose  that $\chi$ is determined mostly by kinematics rather than by specific details of the molecular interactions, and that this property of $\chi$ is largely responsible for the universality observed at high strain rates.

At low temperatures, where changes in the state of glassy configurational disorder can be induced only by externally driven deformation and not by thermal fluctuations, i.e. where $\rho = 0$, there must be a direct relation between the dimensionless strain rate $q$ and the low-temperature effective temperature, say $\chi =\hat\chi(q)$. When the strain rate is much smaller than any relevant relaxation rate in the system, so that $q \ll 1$, then dimensional analysis based on the fact that there is no intrinsic rate comparable to $q/\tau_0$ requires that $\hat\chi(q)$ be equal to a constant, say $\hat\chi_0$.  

In the opposite limit, when $q$ is of the order of unity, we assume that the relation between $q$ and $\hat \chi$ has an Arrhenius form, $q \sim q_0\,\exp\,(-\,A/\hat\chi)$.  Both $\hat\chi$ and $A$ are energies measured in units of $e_Z$.  If $e_Z$ is the only energy scale in the system relevant to configurational rearrangements, then we expect that $A$ is of the order of unity.  Similarly, if $\tau_0^{-1}$ is the only intrinsic rate, then $q_0$ also should be of the order of unity; but here there is an additional uncertainty about how accurately we have estimated $\tau_0$.  We return to these estimates in Sec.~\ref{parameters}.

The relation between strain rate and configurational disorder was discovered in numerical simulations by Haxton and Liu \cite{HAXTON-LIU-07}.  It was discussed further in \cite{JSL-MANNING-TEFF-07}, where it was written in the form of a glasslike relation between a ``viscosity'' $q^{-1}$ and the temperature $\hat\chi$:
\begin{equation}
\label{q-chi}
{1\over q} = {1\over q_0}\,\exp\left[{A\over \hat\chi} + \alpha_{e\!f\!f}(\hat\chi)\right];
\end{equation}
and, in analogy to Vogel-Fulcher-Tamann (VFT), 
\begin{equation}
\label{alphaeff}
\alpha_{e\!f\!f}(\hat\chi)= {\hat\chi_1\over \hat\chi - \hat\chi_0}\,\exp\left[- \,b\,{\hat\chi - \hat\chi_0\over \hat\chi_A - \hat\chi_0}\right].
\end{equation}
Thus, $\hat\chi \to \hat\chi_0$ in the limit of small strain rates, and $\hat\chi \to \infty$ as $q \to q_0$. The exponential cutoff in Eq.(\ref{alphaeff}) is needed in order that the VFT divergence at small $\hat\chi$ transform smoothly to the Arrhenius law at large $\hat\chi$.  Previous calculations have used $b=3$.  Experience with these formulas, as in \cite{JSL-MANNING-TEFF-07}, leads us to conclude that they are more cleanly reliable than their VFT analog for viscosity as a function of ordinary temperature.  (In this connection, see the calculation of the viscosity in  \cite{JSL-11}.) 

The equation of motion for $\chi$ itself is a statement of the first law of thermodynamics; it describes entropy flow through the slow, configurational degrees of freedom into the fast thermal motions of the molecules.  Near steady state, it has the form
\begin{equation}
\label{chidot}
\dot\chi \propto e^{-1/\chi}\left[\Gamma(s,\chi)\,\Bigl(\hat\chi(q) - \chi\Bigr) + \kappa\,\rho(\theta)\,\left(\theta - \chi\right)\right].
\end{equation}
The first term in square brackets on the right-hand side is the rate at which $\chi$ is driven toward $\hat\chi(q)$ by the mechanical noise strength $\Gamma(s,\chi)$.  The second term, proportional to $\rho(\theta)$, is the rate at which thermal fluctuations  drive $\chi$ toward the ambient temperature $\theta$.  $\kappa$ is a dimensionless parameter of the order of unity.  Since $\chi$ is a measure of configurational disorder, we see here explicitly how $\rho(\theta)$ controls the rate of structural aging in undriven systems. The competition between these two terms in Eq.(\ref{chidot}) determines the value of $\chi$; it is close to $\hat\chi(q)$ for large $\Gamma(s,\chi)$, and close to $\theta$  when the system is driven slowly so that $\Gamma(s,\chi)$ is small.  

There is one complication that must be dealt with at this point.  Equation (\ref{chidot}), as written, implies that the steady-state $\chi$ must lie in the interval between $\hat\chi(q)$ and $\theta$.  If we assume that $\hat\chi(q) \approx \hat\chi_0$ is a constant for small enough $q$, then a system initially prepared with $\theta > \hat\chi_0$ would be ``cooled'' to $\chi < \theta$ when driven at a small strain rate.  This behavior seems implausible; so far as we know, it is not seen in other simulations, e.g. \cite{HAXTON-12}, in which $\chi$ is measured directly.  In \cite{JSL-MANNING-TEFF-07}, this problem was corrected by setting $\hat\chi_0 = \theta$ when $\theta$ exceeds $\chi_0$, and by rescaling $\hat\chi_1$ and $\hat\chi_A$ accordingly.  Specifically,
\begin{equation}
\hat\chi_0 = \cases{\chi_0, & for $\theta < \chi_0$,\cr \theta,& for $\theta > \chi_0$};
\end{equation} 
\begin{equation}
\hat\chi_1 = \cases{\chi_1,&  for $\theta < \chi_0$,\cr (\hat\chi_0/\chi_0)\,\chi_1,& for $\theta > \chi_0$};
\end{equation}
and
\begin{equation}
\hat\chi_A = \cases{\chi_A,&  for $\theta < \chi_0$,\cr (\hat\chi_0/\chi_0)\,\chi_A,& for $\theta > \chi_0$}.
\end{equation}

In summary, we use Eq.(\ref{chidot}) in the form
\begin{equation}
\label{chidot2}
\chi = {\Gamma(s,\chi)\,\hat\chi(q) + \kappa\,\rho(\theta)\,\theta\over \Gamma(s,\chi)+ \kappa\,\rho(\theta)},
\end{equation}
along with Eq.(\ref{q-chi}), to determine one relation between $\chi$, $q$ and $s$.  We then use Eq.(\ref{q-sol}) to compute both $q$ and $\chi$ as functions of $s$.  

\begin{figure}[here]
\centering \epsfig{width=.45\textwidth,file=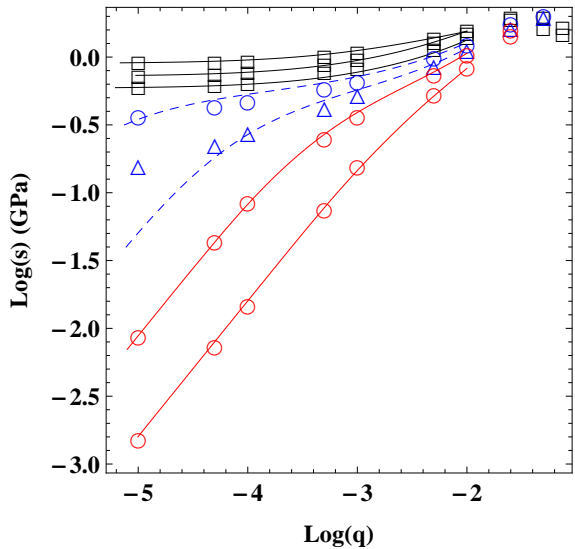} \caption{(Color online) Log-log plot of stress $s$ {\it versus} dimensionless plastic strain rate $q = \tau_0\,\dot\gamma^{pl}$.  The top three solid black theoretical curves and the associated open-square data points are for temperatures $T= 100\,K$, $300\,K$, and $500\,K$.   The middle dashed blue curves are for $T = 700\,K$ (blue circles), and $840\,K$ (blue triangles). The bottom two solid red theoretical curves and open-circle data points are for $T = 940\,K$ and $1100\,K$. Agreement between theory and simulation is good for the low temperatures (black curves) at the top and for the high temperatures (red curves) at the bottom; but it fails at small strain rates for the intermediate temperatures near the glass transition.} \label{GCE1}
\end{figure}

\section{Evaluation of Parameters and Comparison with the GCE Data}
\label{parameters}

In Fig.~\ref{GCE1}, we show a selection from the raw data of GCE. The data sets shown here are characteristic of the eleven such sets that we have used in these analyses. They are stress {\it versus} strain-rate curves for seven different  temperatures, $T= 100\,K,~~300\,K,~~500\,K,~~700\,K,~~ 840\,K,~~940\,K$ and $1100\,K$, reading from top to bottom.  The data points are shown here along with our best-fit theoretical curves. The GCE time constant is $\tau_0 = 0.1$ picoseconds; therefore, the dimensional strain rate is $\dot\gamma^{pl} = 10^{13}\,q\,\,s^{-1}$. We have arbitrarily chosen the upper limit of validity for the STZ theory to be at $q = 10^{-2}$, i.e. at $\dot\gamma^{pl} = 10^{11}\,s^{-1}$.  Beyond this point, the simulations show strain-rate softening -- possibly an indication that the system is liquifying and is no longer consistent with the solidlike STZ model.

Our strategy for choosing parameters has been to look for the simplest possible version of the theory that is consistent with the data.  Accordingly, our starting assumption is that the Eyring-like rate factor is negligible; that is, we set $\theta_E = 0$ in Eq.(\ref{calC}).  We have tried various nonzero values of $\theta_E$,  and have found no overall improvement in the results.  At low temperatures, only large stresses -- above the yield stress -- come into play, reducing the activation barrier via the Gaussian factor in Eq.(\ref{calC}).  Conversely, at high temperatures, the driving stresses may be small, but the activation barriers are small compared to $k_BT$.  The exceptions are at the intermediate temperatures, for small stresses and small strain rates, where the theory breaks down for the more interesting reasons mentioned in Sec.~\ref{intro}. If we choose $\theta_E$ large enough that, just by itself, it substantially increases the viscosity and improves the fit, for example, at $T = 840\,K$ (the lower dashed curve in Fig.~\ref{GCE1}), then we qualitatively ruin the fit at $T = 1100\,K$. Ultimately, the Eyring barrier must be important at smaller strain rates and low temperatures; but we see no need for it here.  

The dimensionless constants $\epsilon_0$ in Eq.(\ref{qeqn}) and $\kappa$ in Eq.(\ref{chidot}) should be of the order of unity. In the absence of better information, we set $\epsilon_0 = \kappa = 1$. Similarly, we set $e_Z/v_0 = 1$ GPa and, in the next paragraph, use the high-temperature viscosity to check that this estimate is reasonable. 

At high temperatures and small stresses, with ${\cal C}(s) \approx \epsilon_0\,v_0/e_Z \approx 1$, Equation (\ref{eta}) tells us that the linear Newtonian viscosity is
\begin{equation}
\label{viscosity}
\eta_N = \left[2+ \rho(\theta)\right]\,{\theta\,e^{1/\theta}\over 2\,\rho(\theta)}\approx {3\over 2}\,\theta\,e^{1/\theta}.
\end{equation}
The last approximation is valid for temperatures high enough that $\rho \sim 1$.  At $T = 1100\,K$ (the bottom curve in Fig.~\ref{GCE1}), the GCE result is $\eta_N =150$. Then, Eq.(\ref{viscosity}) implies that  $\theta = 0.15$, and therefore that $T_Z = e_Z/k_B \approx 7000\,K$.  This means that $v_0/e_Z \approx 10\,v_0$, with $v_0$ measured in cubic nanometers.  Thus, a length scale of about half a nanometer gives us $e_Z/v_0 \approx 1$ GPa as in the preceding paragraph.  

Now consider low-temperature situations for which the GCE data appear to indicate well defined yield stresses $s_0(\theta)$, implying that $\rho(\theta) = 0$ to within the accuracy of the measurements.  The temperatures for which this is true are $T = 100\,K,~~300\,K,~~500\,K$ and $600\,K$.  In this regime, ${\cal T}(s,\chi) \approx 1$; thus, in the limit of small $q$, Eq.(\ref{qeqn2}) becomes
\begin{equation}
\label{q-sigma-approx}
q \approx 2\,e^{-\,1/\chi_0}\left(1 - {s_0\over s}\right),
\end{equation}
which we have used to make a first estimate of $\chi_0$.  We then have found that we can fit all four of these data sets, for the whole range of strain rates $10^{-5}< q < 10^{-2}$, by using the observed values of $s_0(\theta)$, and by choosing $\chi_0 = 0.1$, $\chi_1 = \chi_A = 0.2$, $A = 1.3$, $q_0 = 5$, and $b = 3$ in Eqs.(\ref{q-chi}) and (\ref{alphaeff}).  We also have chosen the Bagnold temperature defined following Eq.(\ref{calC}) to be $T_B =100\,K$.  With this value of $T_B$, the Bagnold term is negligible for all but the lowest temperatures and largest stresses; but it does seem to be relevant in that regime. We have no reason to believe that this set of temperature-independent parameters is unique or optimal.   

\begin{figure}[here]
\centering \epsfig{width=.45\textwidth,file=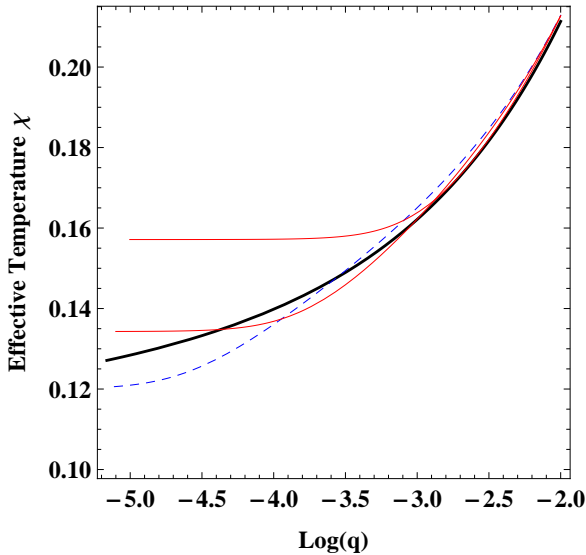} \caption{(Color online) Log-linear plot of the theoretical effective temperature $\chi$ as a function the dimensionless strain rate $q$ for temperatures $T= 100\,K$ through $600\,K$ (dark black line), $840\,K$ (dashed blue line), and $940\,K$ and $1100\,K$ (thin red lines, from bottom to top).} \label{GCE2}
\end{figure}

The only remaining parameters to be chosen are the quantities $\rho(\theta)$ and $s_0(\theta)$ for temperatures near and above the glass transition.  Rather than trying to predict these quantities theoretically, we have ``measured'' them by fitting the GCE data.  The results are shown below in Figs.\ref{GCE3} and \ref{GCE5}, and are discussed in Sec.~\ref{analysis}.  

The theoretical fits to the data in Fig.~\ref{GCE1} are accurate for both the lower and higher temperatures, indicated by solid black and red curves respectively.  However, there are significant discrepancies at small strain rates for the intermediate temperatures $T = 700\,K$ and $840\,K$, shown by the dashed blue lines. (There is a similar discrepancy for $T = 800\,K$, not shown in the figure.)   At $T = 840\,K$, for example, the theoretical stress starts to fall below the observed values when $q$ becomes less than about $10^{-4}$; and the system switches over to linear viscosity at smaller strain rates.  The multi-species STZ theory proposed in \cite{BL-11,JSL-11} predicts that the same thing will happen whenever $\rho$ is  nonzero, but that the transition will occur at a smaller $q$ and, therefore, at a larger viscosity.  

In Fig.~\ref{GCE2}, we show four examples of our theoretical $\chi$'s as functions of strain rate. All four converge to the same large-$q$ behavior shown in Eq.(\ref{q-chi}).  The heavy black curve is the low-temperature function, $\chi= \hat\chi(q)$, valid for all cases in the range $100\,K < T < 600\,K$ where $\rho = 0$.  It decreases logarithmically at small $q$ toward $\chi_0 = 0.1$.  The two red curves, for $T= 940\,K$ and $1100\,K$, where $\rho$ is large, fall quickly to their corresponding values of $\chi = \theta = T/T_Z$ at small $q$.  The blue dashed curve is for $T = 840\,K$, where $\rho \cong 0.1$.  This curve falls substantially below the $\rho = 0$ curve before levelling off at $T/T_Z = 0.12$, again an indication that the small-$q$ behavior is incorrect.  The theory predicts too small a viscosity in this case; therefore, it predicts too small a value of $\Gamma(s,\chi)$ relative to $\rho$ in Eq.(\ref{chidot2}), and thus predicts too fast an approach to the thermal limit.  Note that the largest values of $\chi$ shown in this figure are still small enough that the STZ density $\sim  \exp\,(-1/\chi)$ remains small.  

\begin{figure}[here]
\centering \epsfig{width=.45\textwidth,file=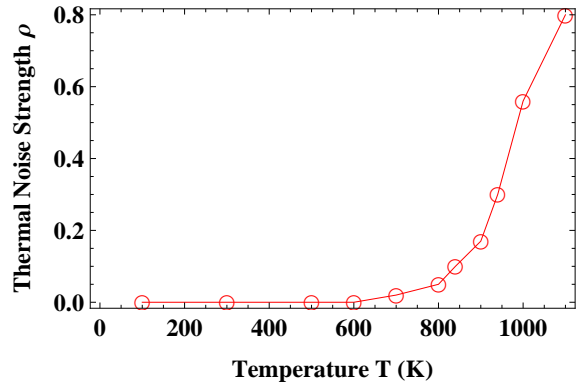} \caption{(Color online) Thermal noise strength $\rho$ (open red circles) as a function of temperature $T$.} \label{GCE3}
\end{figure}

\section{Data Analysis}
\label{analysis}

In principle, an ideal glass transition occurs in an undriven, equilibrated, amorphous system at a Kauzmann temperature $T_0$, below which the viscosity is infinite and, above which, the yield stress vanishes.  The GCE scaling analysis implies that something like this transition is occurring in this rapidly deforming system.  

To begin to understand this situation, we show in Fig.~\ref{GCE3} our ``measured'' values of the thermal noise strength $\rho$ for the eleven different temperatures included in the GCE data.  This function behaves more or less as expected, falling from a value of the order of unity at large $T$ to become unmeasurably small below the glass transition. The most notable feature of this curve is the smooth transition to zero in the region $600\,K < T < 840\,K$.  For these temperatures, we have optimized the fits to be accurate at the higher strain rates, as shown in Fig.~\ref{GCE1} by the dashed curves for $T = 700\,K$ and $840\,K$.  The single-species STZ theory should be valid at large $q$.  Therefore, these measurements of $\rho$ must be taken seriously despite the fact that they appear to be inconsistent with VFT formulas for $\rho$ of the kind developed in \cite{Langer-PRE-06,Langer-PRL-06}.

\begin{figure}[here]
\centering \epsfig{width=.45\textwidth,file=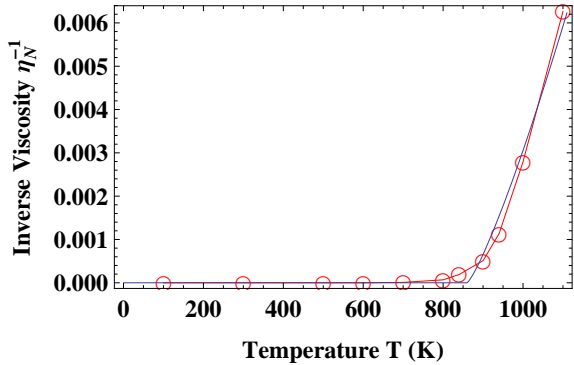} \caption{(Color online) Inverse Newtonian viscosity $\eta_N^{-1}$ (open red circles) as a function of temperature $T$.  Also shown (blue solid line) is the GCE fit to this curve given in Eq.(\ref{etaNfit}).} \label{GCE4}
\end{figure}

The glass transition emerges more clearly if, instead of plotting $\rho$ itself, we use the $\rho$-dependent formula for $\eta_N$ in Eq.(\ref{viscosity}) to plot the inverse Newtonian viscosity  $\eta_N^{-1}$ as a function of temperature.  This is the function that GCE suggest can be fit by
\begin{equation}
\label{etaNfit}
\eta_N^{-1}(T) = \cases{{\rm constant}\times (T - T_0)^{\alpha} & for $T > T_0$\cr 0 &otherwise},
\end{equation}
where $T_0 = 860\,K$ and $\alpha = 1.23$.  Our result and the comparison with the GCE fit are shown in Fig.~\ref{GCE4}.  Here, the fact that we are underestimating the viscosities for temperatures less than $T_0$ does not make much difference because $\eta_N^{-1}$ is already very small in that region.  

\begin{figure}[here]
\centering \epsfig{width=.45\textwidth,file=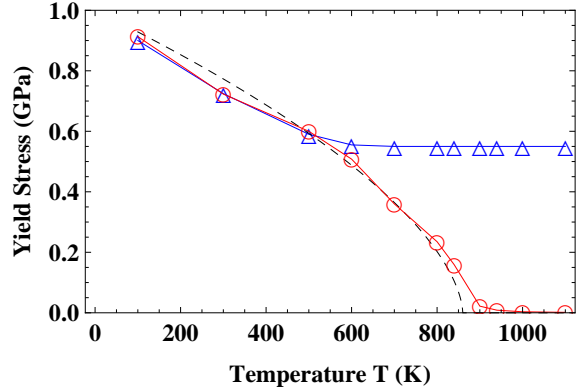} \caption{(Color online) Apparent yield stress from GCE data (open red circles) and the GCE fit to this data (dashed black curve) given in Eq.(\ref{sigmayfit}).  The open blue triangles show the measured values of the stress $s_0$. } \label{GCE5}
\end{figure}

According to the STZ theory, there is no yield stress in the limit of zero strain rate when $\rho$ is nonzero;  the system must have a finite linear viscosity at small enough stresses and strain rates. Nevertheless, the theory predicts that, at very small $\rho$, the system will appear to have a yield stress so long as the strain rate is not too small.  Therefore, we can plot this apparent yield stress by plotting the observed stress at the smallest observed strain rates, as a function of temperature.  The GCE data for $q = 10^{-5}$ is shown in Fig.~\ref{GCE5}, along with the GCE fit to this apparent yielding curve: 
\begin{equation}
\label{sigmayfit}
s_y^{app} = \cases{{\rm constant} \times (T_0 - T)^{\beta} &for $T < T_0$\cr 0&otherwise},
\end{equation}
where $\beta = 0.6$.  The yield stress predicted by the STZ theory would be given by the blue triangles in this figure for temperatures up to about $600\,K$ (for which $\rho = 0$), and then would drop abruptly to zero above that point.  It would be interesting to see whether something closer to this behavior would appear if the GCE data were extended to smaller strain rates. We also show our measured values of the stress $s_0$ in Fig.~\ref{GCE5}.  In accord with the preceding remarks, $s_0$ coincides with the yield stress at temperatures up to $600\,K$.  At higher temperatures, as found in \cite{JSL-MANNING-TEFF-07}, $s_0$ levels off at about $0.55$ GPa.  This simple behavior is determined by the large-$q$ dependence of the high temperature curves in Fig.~\ref{GCE1}, where the onset of nonlinearity is sensitive to $s_0$. 

\begin{figure}
\centering \epsfig{width=.45\textwidth,file=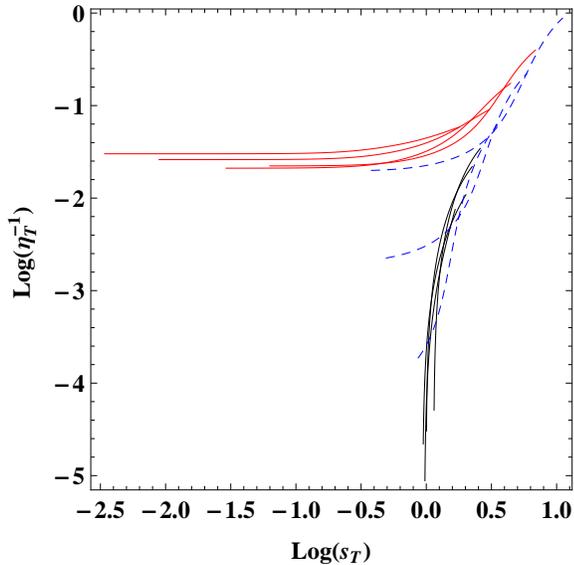} \caption{(Color online) Scaled inverse viscosity $\eta_T^{-1}$ as a function of the scaled stress $s_T$, computed from the STZ theory using all eleven GCE temperatures.  The color scheme is analogous to that used in Fig.~\ref{GCE1}. The four solid red curves that collapse approximately to a single scaled inverse viscosity are for high temperatures in the range $900\,K - 1100\,K$.  The four solid black curves that collapse approximately to a single scaled yield stress are for temperatures in the range $100\,K - 600\,K$. The three dashed blue curves, which deviate from the scaling pattern, are for temperatures $700\,K,~ 800\,K$ and $840\,K$, reading from bottom to top. } \label{GCE6}
\end{figure}

The power-law approximations shown in Eqs.~(\ref{etaNfit}) and (\ref{sigmayfit}) are similar to those used by Olssen and Teitel \cite{OLSSEN-TEITEL-07} in their analysis of a zero-temperature, jamming-unjamming transition controlled by the volume fraction of a model granular material.  The GCE analog for the temperature controlled situation is  
\begin{equation}
\label{scaling}
{1\over \eta_T} = {1\over \eta(s)}\,\left|{T\over T_0}- 1\right|^{-\alpha},~~~s_T = s\,\left|{T\over T_0}- 1\right|^{-\beta},
\end{equation}
where $\eta(s) = s/q$ (not the small-$q$ Newtonian viscosity).  A  graph of $\eta_T^{-1}$ as a function of $s_T$ produced the data collapse shown in GCE Fig. 2.    Our STZ version of that figure is shown here in Fig.~\ref{GCE6}, where we have included the data for all eleven of the GCE temperatures. 

Note that the high-temperature (upper) branch in Fig.~\ref{GCE6} has become nearly a single horizontal line at the constant inverse viscosity  $\eta_T^{-1}$ given in Eq.(\ref{scaling}).  These lines do not lie exactly on top of each other in our graph because, as seen in Figs.~\ref{GCE4} and \ref{GCE5}, the power-law fits in Eqs.~(\ref{etaNfit}) and (\ref{sigmayfit}) are quantitatively inaccurate. The low-temperature (lower) branch is not quite so clean, because $\eta_N^{-1}$ is small but nonzero for temperatures near $T_0$ as shown in Fig.~\ref{GCE4}.  The resulting failure of the scaling analysis appears in the three dashed curves in Fig.~\ref{GCE6} for the temperatures $T = 700\,K,~ 800\,K$ and $840\,K$, where the low-temperature behavior crosses over to an apparent linear viscosity with decreasing strain rate. This effect would be even more pronounced if, as predicted, the actual yield stress drops to zero above  $600\,K$.  Apparently, the data collapse seen in GCE does not indicate an exact scaling relation or the presence of some diverging length scale.  Nevertheless, it remains an interesting qualitative feature of the data.     

\section{Concluding Remarks}
\label{conclusions}

The analysis presented here makes it appear that the STZ theory of amorphous plasticity is realistic up to strain rates of about $10^{-2}/\tau_0$, i.e. up to about one percent of the underlying molecular relaxation rate.  Both the GCE simulations and our theoretical results exhibit an apparently simple scaling behavior across three decades of strain rates below that limit.  Our STZ analysis implies, at least in a first approximation, that effective-temperature thermodynamics, rather than material-specific molecular interactions, control the system dynamics at high strain rates. 

As discussed in Sec.~\ref{intro}, these STZ-based results are technically beyond the range of the mode-coupling theory \cite{GOTZE-91,GOTZE-92,BRADER-et-al-09}, because they extend down to temperatures well below the glass transition where plastic flow is governed by nonperturbative activation mechanisms.  However, see \cite{VOIGTMANN-11} for a counter argument.  Similarly, we do not see how our results can be reproduced in the framework of soft glassy rheology (SGR) \cite{SGR-97,SOLLICH-98}, because the STZ and SGR theories are so different kinematically, and because our present analysis is based so strongly on effective-temperature dynamics.  Again, however, we point to \cite{SOLLICH-CATES-12}, which describes progress in reformulating SGR in a thermodynamic framework analogous to that of \cite{BLII-III-09}.  It will be interesting to see whether the results presented here do, in fact, distinguish the STZ theory from either of these two other competing points of view. 

Our analysis in this paper is based on several, fundamental hypotheses that are potentially falsifiable, and which require further attention.  First, in order to check the thermodynamic hypothesis, we need to test our theoretical estimate of the effective temperature $\chi$ by measuring it directly.  Computational measurements of $\chi$ in \cite{HAXTON-LIU-07} served as the basis for the theory developed in \cite{JSL-MANNING-TEFF-07}, which led to the equations of motion for $\chi$ presented here in Sec.~\ref{chi}. Analogous measurements of an effective temperature for a thermalized hard-sphere system have been reported recently by Haxton \cite{HAXTON-12}; and an STZ analysis of Haxton's results has revealed interesting insights \cite{LIEOU-JSL-12}.  So far, we have not had a similar consistency check for our  analysis of the GCE data.  

Second, our qualitative explanation of the discrepancies between the GCE data and single-species STZ theory is based on the multi-species reinterpretation of glassy viscosity proposed in \cite{JSL-11}.  We need now to develop this reinterpretation into a quantitative theory, and test whether that theory corrects the discrepancies.  This is our most unconventional effort.  We have used large-strain-rate data, instead of the conventional small-strain-rate viscosity, to evaluate the thermal noise strength $\rho(\theta)$.  This procedure produces unexpectedly large values of $\rho(\theta)$ at temperatures near $T_0$, as seen in Fig.~\ref{GCE3}.  The argument in  \cite{JSL-11} tells us that, because of the statistically inevitable presence of ``slow'' STZ's, the viscosity increases more rapidly with decreasing temperature than would be predicted simply by the $\theta$ dependence of $\rho$.  This argument, when applied to undriven systems near their glass temperatures, accounts for Stokes-Einstein violations and even for the stretched-exponential nature of various relaxation functions. The question is: Does it also account quantitatively for what look like anomalously large viscosities in the GCE data?  

A third class of questions arises naturally within the context of the first two. Our analysis indicates that the model of a metallic glass used in the GCE simulations undergoes some kind of glass transition at $T_0 \cong 860\,K$.  As $T$ decreases below $T_0$, the configurational noise strength $\rho$ decreases smoothly and becomes unmeasurably small below about $600\,K$. As this happens, the system changes from a viscous, glass-forming liquid to a solidlike glass with an apparently well defined yield stress. At nonzero, steady, shear rates, this transition becomes smoother; the liquidlike behavior persists down to arbitrarily low temperatures with continuous increase in the stress level.    

Finally, an obvious question from a first-principles point of view is how the STZ transitions appear in the form of  changes in the underlying atomic structure.  Reference  \cite{IWASHITA-EGAMI-12}, a sequel to the GCE simulation \cite{GCE-10}, is a recent example of work along these lines. It is shown there that, during steady-state flow, the macroscopic Maxwell relaxation time $\eta/G_{\infty}$ (where $G_{\infty}$ is the instantaneous shear modulus) is essentially the same as the time required for a localized topological change to occur in which an atom gains and/or loses one of its nearest neighbors.  We expect that further investigations of this kind will lead to a clearer microscopic picture of the STZ transition mechanism.

\begin{acknowledgments}

This research has been sponsored by the U.S. Department of Energy, Basic Energy Sciences, Materials Sciences and Engineering Division. JSL thanks Michael Cates for especially helpful discussions. 

\end{acknowledgments}

\end{document}